# Silicon microcavity arrays with open access and a finesse of half a million

Georg Wachter[1,2], Stefan Kuhn*[,1], Stefan Minniberger[1], Cameron Salter[1], Peter Asenbaum[1], James Millen[1,+], Michael Schneider[3], Johannes Schalko[3], Ulrich Schmid[3], André Felgner[4], Dorothee Hüser[4], Markus Arndt[1], and Michael Trupke*[,1,2]

[1]*Faculty of Physics, University of Vienna, VCQ, Boltzmanngasse 5, 1090 Vienna, Austria*
[2]*Institute for Atomic and Subatomic Physics, TU Wien, VCQ, Stadionallee 2, 1020 Vienna, Austria*
[3]*Institute for Sensor and Actuator Systems, TU Wien, 1040 Vienna, Austria*
[4]*Physikalisch-Technische Bundesanstalt, Bundesallee 100, D-38116 Braunschweig*
[+]*Current address: Department of Physics, King's College London, Strand, London WC2R 2LS United Kingdom*

**Optical resonators are increasingly important tools in science and technology. Their applications range from laser physics, atomic clocks, molecular spectroscopy, and single-photon generation to the detection, trapping and cooling of atoms or nano-scale objects**[1–5]**. Many of these applications benefit from strong mode confinement and high optical quality factors, making small mirrors of high surface-quality desirable. Building such devices in silicon yields ultra-low absorption at telecom wavelengths and enables integration of micro-structures with mechanical, electrical and other functionalities**[6,7]**. Here, we push optical resonator technology**[8] **to new limits by fabricating lithographically aligned silicon mirrors with ultra-smooth surfaces, small and well-controlled radii of curvature, ultra-low loss and high reflectivity. We build large arrays of microcavities with finesse greater than F = 500,000 and a mode volume of 330 femtoliters at wavelengths near 1550 nm. Such high-quality micro-mirrors open up a new regime of optics and enable unprecedented explorations of strong coupling between light and matter.**

Scalable photonic technologies require optical devices which are compact, can be mass-produced and integrated while maintaining the best performance. Optical cavities with high field enhancement, low mode volume and narrow linewidths are of particular importance for sideband-resolved light-matter interactions[4,9], in transition-selective photon sources[10] or in nanoparticle cooling experiments[9,11–13]. In a two-mirror Fabry-Pérot (FP) geometry, these quantities are determined by the quality and shape of the mirror substrate, its coating and the mirror separation. Microscopic surface roughness and deviations from a perfect parabolic shape cause scattering of cavity light into higher-order modes and free-space, resulting in photon loss[14]. In addition, the cavity performance depends on the mirror alignment; displacements or tilts can lead to clipping losses. For many applications, such as cooling or detection of nanoparticles, it is also important to have free access to the optical mode. This is a geometric challenge in a micro-design with strict alignment requirements.

To precisely manufacture low-loss silicon mirrors with well controlled curvature we use a CMOS-compatible etching process, as depicted in Fig. 1a. The mirror geometry, including its curvature and depth, are carefully engineered by the choice of parameters in a two-step dry-etching process. We achieve a surface quality approaching the atomic limit by furthermore applying a series of oxidation and HF-etching steps[15] (see Methods and Supplementary Information).

The mirror chips are coated with high-reflectivity dielectric layers (T < 5 ppm) on their micro-structured side and with antireflection-coating with a reflectivity of $\rho = 0.1$ % on the other side. Figure 1b shows a chip containing 100 micro-mirrors, all with different curvatures to demonstrate the versatility of the method. The same coatings are applied to a set of planar silicon chips. These chips allow us to realise two different cavity geometries: Plano-concave (PC) cavity arrays (see Fig.1d) with a minimal mode volume down to 330 femtoliters and symmetric concave-concave (CC) cavities with an alignment spacer (see Fig. 1e) to optimise the light-matter coupling by stronger mode confinement at the cavity centre.

*Correspondence and requests for materials should be addressed to stefan.kuhn@univie.ac.at or michael.trupke@univie.ac.at

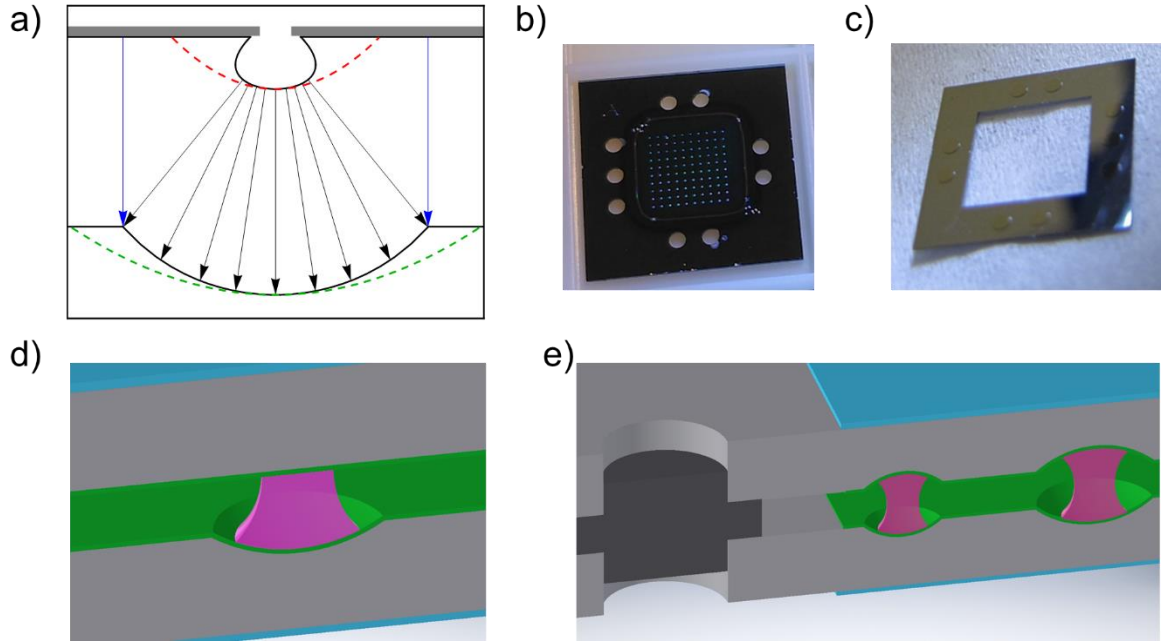

**Figure 1:** *Fabrication and assembly of the micromirrors. a) A photoresist mask (grey) is used to expose well-defined circular apertures on a silicon wafer. Plasma etching then creates concave shapes which still deviate from the desired profile (dashed red line). A second, mask-less etch step enlarges the central portion, thereby approaching an ideal parabola (green). We use a variety of aperture sizes in the initial etching mask to create a range of mirror radii of curvature. The final opening size is defined by the concurrent vertical etch (blue arrows) of the flat silicon surface. b) A typical mirror chip (dark grey) contains 100 micromirrors of different sizes and radii, with a square pitch of 500 micrometres. It is coated with a high-reflectivity Bragg layer, visible as a dark-green square. Circular holes are etched into the frame to enable rigid mechanical alignment. c) The silicon spacer with cylindrical alignment pillars matches the holes in the mirror chips. d) Schematic of the plano-concave mirror design. A concave micro-mirror in silicon (grey) is combined with a plane mirror. Both are coated with the same high reflectivity multilayer Bragg stack (green). Their back sides are antireflection-coated (light blue). The two mirrors define the boundary conditions for the optical mode (purple). e) Schematic of the symmetric silicon microcavity array with concave mirrors. The two devices are rigidly aligned and separated by an interlocking silicon spacer.*

For both configurations we determine the properties of the microcavity array by measuring the infrared transmission through a single mirror pair. By scanning a laser (Toptica CTL) between $1520 - 1630$ nm we measure the free spectral range $FSR$ to determine the cavity length $L = c/2\,FSR$, where $c$ is the speed of light (Fig. 2a). In the PC design, $L$ ranges from 17 µm to 24 µm. In the CC design, $L = 140 - 160$ µm and can be controlled by choice of the thickness of the spacer.

From the frequency separation of higher transverse cavity modes (see Fig. 2a) we find the mirror radii to range between 123 µm and 289 µm (see Methods). To further assess the mirror quality we measure the finesse $F$ of the individual cavities using both cavity ring down and side-band modulation spectroscopy. In the first scheme, a fibre-based acousto-optic modulator (f-AOM) is used to rapidly switch off the pump laser field while the cavity is locked close to resonance. The photon lifetime or cavity decay can then be directly monitored on a fast photodiode (150 MHz bandwidth), as shown in Fig. 2b. Alternatively, we use a fibre-based electro-optic modulator (f-EOM) to imprint well-defined frequency sidebands onto the carrier beam which is then scanned across the cavity resonance. Using the frequency separation of the sidebands we can directly determine the cavity linewidth in a fit to the transmission spectrum, as shown in the inset of Fig. 2b.

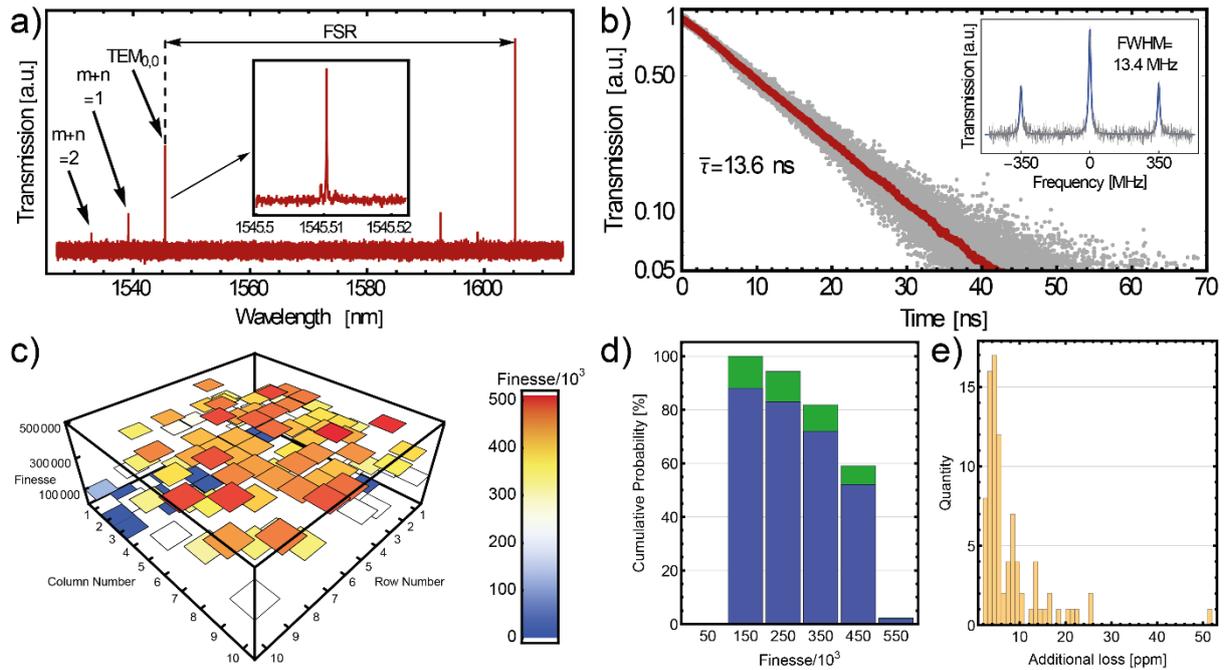

**Figure 2:** *Performance of the plano-concave microcavities. a) Spectrum of a microcavity showing the free spectral range (FSR) as well as higher-index modes which are used to determine the radius of curvature (see methods). The inset shows the lowest-order mode. b) A cavity ring-down measurement yields a lifetime of (13.6±0.3) ns. The inset shows a single scan over the cavity resonance recorded with an applied laser frequency sideband-modulation of 350 MHz. The frequency scale given by the resulting sidebands yields a linewidth of κ/2π=(6.77±0.86) MHz. With the FSR=7.23 THz determined as in a), we measure a finesse of F=(5.3±0.6)×10⁵ from the linewidth scan and F=(6.2±0.2)×10⁵ for the ring-down measurement, respectively. The uncertainties are the standard deviations over 75 and 992 measurements for the spectral and ringdown measurements, respectively. c) Distribution of the finesse over an array of 100 microcavities. d) Cumulative distribution of the finesse values. A resonance was found for 88 out of 100 cavities. The bars show the probability for all cavities (blue bars) and for functioning cavities (green bars). e) Distribution of losses that we find in addition to the specified 5 ppm transmission of the mirror coatings.*

Figure 2c shows the finesse of all 100 individual cavities on a PC array where the etch mask radius increases from 6.2 µm to 26 µm in steps of 200 nm from row 1, column 1 to row 10, column 10. Figures 2d-e show that a high level of performance is achieved for the entire array. Our measurements yield a maximum finesse of $F = (5.0 \pm 0.1) \times 10^5$ for $L = 16.8$ µm and $R = 166$ µm. This mirror was formed with an initial mask opening radius of 9.8 µm. This corresponds to a mode volume of only $V = 330$ fL or as little as $88\lambda^3$, while achieving an optical quality factor of $Q = (1.1 \pm 0.05) \times 10^7$. These values compare favourably with those achieved for micro-pillar structures used to generate single photons from integrated quantum dots[16].

We have further created rigidly assembled arrays of concave mirror pairs. Accurate and robust alignment of the cavities is achieved by applying lithographically precise micromachining to form through-etched alignment holes into the mirror chips. In a separate fabrication run, spacer chips with a thickness of 100 µm are created with 20 µm high micro-pillars to fit into these alignment holes. This provides precise, lithographically defined alignment and controlled spacing between the mirror chips (see Methods). The finesse for a selection of CC cavities is plotted in Fig. 3a, with a peak value of $F = (4.0 \pm 0.1) \times 10^5$ for a cavity with $R = 201$ µm and $L = 150$ µm. These values correspond to $Q = 7.9 \times 10^7$. The fact that such high finesse could be achieved at a value of $L/R = 0.75$ indicates that the micromachined cavity pre-alignment avoids clipping losses of the optical mode at the mirror edges.

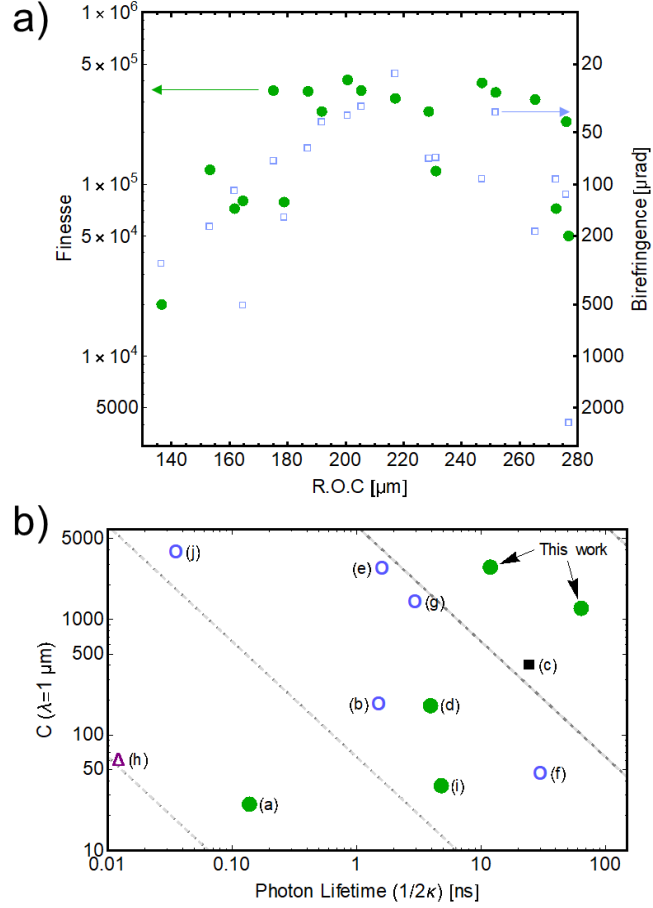

**Figure 3:** *Performance of symmetric microcavities and state of the art. a) Finesse (red circles) and birefringence (blue squares) of all measured CC cavities. b) Comparison of cooperativity C and photon lifetime for state-of-the-art microcavities ((a)[17], (b)[18], (c)[19], (d)[15], (e)[20], (f)[21], (g)[22], (h)[23], (i)[9], and (j)[24]). Dashed lines indicate $C(\tau)$ for constant finesse $F=10^p$ with p=(1,2,3,4) and varying L=R in a symmetric cavity. Light blue circles: micro-machined $SiO_2$. Black square: macroscopic $SiO_2$ mirrors. Purple triangle: Buckled-dome cavities. Green dots: Silicon micromirrors, including the best results from this work, as indicated, for PC (left dot) and CC (right dot) configurations.*

A remarkable feature of our optical resonators is their low birefringence, which is comparable to the cavity linewidth for all tested cavities. A useful measure for birefringence is the differential phase shift $\delta\phi$ per round-trip accumulated between orthogonal polarizations of the light field, and is given by the ratio of the frequency splitting $\delta f$ between the polarization-dependent modes to the $FSR$: $\delta\varphi = 2\pi \cdot \delta f/FSR$. It correlates with the finesse, as shown in Fig. 3a. The high performance achieved in both parameters constrains possible imperfections to a small, long-range surface roughness with a spatial frequency on the order of the mode waist. For our best CC cavities the phase shift corresponds to less than $\delta\varphi = 23$ µrad per round-trip. This value is comparable to the best values reported for laser machined mirror cavities[21,22]. It suggests that $R$ is nearly constant in all polar directions, which is consistent with isotropic etching and precise alignment.

A finesse of $F > 5 \times 10^5$, as found for our PC cavities, is very close to the value of $F = 6.3 \times 10^5$ expected from the target transmission in the coating process. Losses due to microscopic surface roughness or shape deformations must therefore be smaller than 2.8 ppm for our best cavities (see Fig. 2e and Methods). This small value must be taken as an upper limit since it neglects residual absorption by the reflective coating and loss induced by the flat mirror. For a large fraction of CC cavities we find a comparably low loss value, which corroborates our assumption that the microfabrication precision fulfils its purpose and ensures high alignment quality. In the future a finesse of $F=10^6$ thus appears possible using absorption-limited coatings.

The outstanding performance of our microcavities will be of great utility for a wide range of applications. Their high finesse, strong field confinement and narrow linewidth will be important for manipulating the internal states of effective two-level systems – such as atoms or solid-state emitters – or the motional states of optomechanical systems such as levitated nanoparticles and membranes[3,4,8].

In many applications the relevant figure of merit is the cooperativity parameter $C = g^2/2\kappa\gamma$ which compares the light-matter coupling frequency $g$ – in atoms: the Rabi frequency – to cavity and matter-related damping terms $\kappa$ and $\gamma$, respectively. Large values of $C$ are desirable for efficient energy exchange between the cavity and the particles. Regardless of the specific system it can be maximised by the cavity parameters as

$$C = \frac{3\lambda^2}{\pi^3}\frac{F}{w_C^2}, \text{ with the mode waist } w_C = \sqrt{\frac{N_C}{\sqrt{LR(N_C-L/R)}}}.$$

This expression is valid for PC cavities with $N_C = 1$ and for symmetric CC cavities with $N_C = 2$. High cooperativity thus requires a high finesse $F$ and strong mode confinement. For open-access, FP cavities using micromirrors, a small waist combined with a high finesse is however only achievable for short cavities due to diffraction-induced clipping losses[25]. This limitation results in an inherent trade-off between photon lifetime and cooperativity.

In Fig. 3b, we plot $C$ against the photon lifetime $\tau = 1/2\kappa$ for a variety of microcavity systems, where we have selected FP resonators with a length below 1 mm. We have included micro-machined and macroscopic $SiO_2$ mirrors, buckled-dome cavities, and silicon micromirrors. For the purpose of comparison, we have re-calculated all cooperativity values for $\lambda = 1$ µm. Due to the strong mirror curvature and high optical finesse our cavities simultaneously attain extremely large cooperativity and photon lifetimes of several tens of nanoseconds, corresponding to MHz-range linewidths, as required for many of the applications mentioned above.

Higher cooperativity $C$ can be achieved by further reducing the mirrors' radii-of-curvature. This will require adapting the etching parameters, but previous measurements indicate that a reduction by an order of magnitude is realistic[7]. A further increase of $C$ is possible by stretching the cavity length $L$ for longer photon lifetimes. A further increase of $C$ is possible by combining a micro-mirror of $R = 169$ µm with a macroscopically curved substrate, e.g. with $R = 50$ mm, both coated for $F = 5 \times 10^5$. This device will enable a cooperativity of 2.8×10³ with a linewidth of only $\kappa/\pi = 6$ kHz.

In summary, the micro-mirrors and open-access microcavity structures presented herein combine extremely low losses and strong mode confinement with scalable micromachining methods and precise alignment. These features will be of benefit for fundamental science and applied quantum technologies[9,26,27]. Future spacer designs can integrate quantum emitters, light guides, detector structures or optomechanical systems within the cavity frame, enabling precise overlap of the cavity field with the desired system to fully exploit the high performance of the microcavity arrays.

## Methods

*Fabrication:* The micro-mirrors are etched into a single-crystal silicon wafer with (100) cut and weak n-doping to about $50\ \Omega$/cm. The etch masks are formed by adding three layers of photoresist (AZ6624), with 3 µm thickness each. The etching is performed in an $SF_6$ plasma at a flow rate of 100 sccm, a temperature of 30°C, an inductively coupled plasma power of 2 kW and a table power of 15 W. The masked etch step lasts for 320 s. The photoresist is then removed in ultrapure acetone and the entire wafer is etched for another 45 minutes using the same recipe. We derive a rate of 4.2 µm/min for the masked etch. The mask-less etch rate is reduced to 0.9 µm/min due to the increased consumption of plasma by the far greater exposed silicon surface. A smoothing procedure using wet oxidation to a thickness of 2 µm, followed by oxide removal using hydrofluoric acid, is then repeated twice to improve the surface quality of the mirror substrate. Bosch etching was used to create the circular holes in the chips, and to separate the devices in a single step. The spacer chips and the alignment pillars on them were created by two further Bosch processes. To facilitate the final assembly, the chips were subjected to a 30 s isotropic etch. This step rounds off the edges of the pillars, and results in a reduction of the pillar radius by 0.5 µm. Taking into account this reduction, and the intrinsic precision of lithographic processing, we expect the relative positional accuracy of two opposing mirrors to be better than 1 µm.

Lastly, the microchips were secured in an aluminium mount for coating. A Bragg mirror coating, consisting of 36 alternating $\lambda/4$ –layers of silicon dioxide ($n = 1.45$) and tantalum pentoxide ($n = 2.04$) was applied to the front of the chips. The back side was broadband anti-reflection coated with an optimized five-layer coating using the same materials.

*Cavity waist:* The waist of the optical mode in a Fabry-Pérot type resonator[28] depends on the radius of curvature of both mirrors $R_1$ and $R_2$ and their separation $L$

$$w_C = \sqrt{\frac{\lambda}{\pi}\sqrt{\frac{L(R_1-L)(R_2-L)(R_1+R_2-L)}{(R_1+R_2-2L)^2}}}.$$

This expression simplifies for a symmetric ($R_1 = R_2$) cavity. The beam waist increases with distance from the focal point. On the surface of a curved mirror it is

$$w_M = w_C\sqrt{1+\left(\frac{\lambda L/N_C}{\pi w_C^2}\right)^2},$$

where $N_C$=1 for a PC geometry and $N_C$=2 for a CC cavity geometry. The beam divergence limits the maximal possible finesse, due to the finite mirror size.

*Radius of curvature:* The radius of curvature of a mirror in a Fabry-Pérot cavity can be determined by measuring the frequency spacing of higher-order modes (see Fig. 3 a). The frequency *f* of a mode with longitudinal index *l* and transverse indices (*m, n*) is given by

$$f(l,m+n) = \frac{c}{2L}\left[l+\frac{1+m+n}{\pi}\cos^{-1}\left(\sqrt{1-\frac{L}{R_1}}\sqrt{1-\frac{L}{R_2}}\right)\right].$$

For PC and symmetric CC cavities, the frequency difference $\chi$ between the *m+n* transverse mode and the fundamental mode can be divided by the free spectral range to yield

$$\chi = \frac{f(l,m+n)-f(l,0)}{FSR} = \frac{m+n}{\pi}\cos^{-1}\left(1-\frac{L}{R}\right)^{N_C/2}.$$

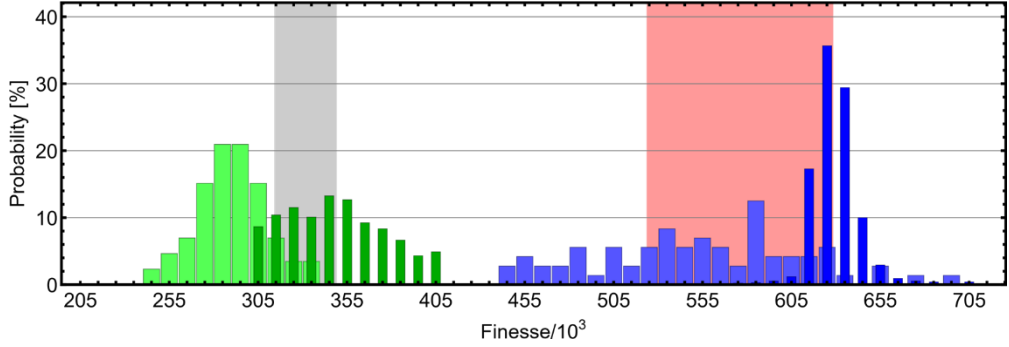

**Figure 4:** *Distribution of observed finesse values for a single, plano-concave microcavity. Finesse distributions calculated from linewidth measurements at 1604 nm (light green, broad bars) and 1545 nm (light blue, broad bars) are compared to ring-down measurements at the same wavelengths (dark green and dark blue narrow bars). The grey and light red bands show the range of finesse values between the design transmission values (9 ppm and 5 ppm) and the maximum loss due to coating imperfections (1 ppm additional scattering loss, both wavelengths).*

The radius of curvature is then given by

$$R = \frac{L}{1 + \left(\cos\frac{\pi\chi}{m+n}\right)^{2/N_C}}.$$

*Measurement of finesse:* The finesse of all cavities is determined first by measuring their linewidth. Since this method may underestimate *F* as laser noise or mechanical noise can increase the measured linewidth, we also measured the photon lifetime using cavity ring-down for selected resonators. This technique may over-estimate the finesse, since electronic response functions will be convolved with the optical signal. In Fig. 4, we show an example measurement of one PC cavity. We compare the statistics of both methods to the properties of the coating as specified by the manufacturer, taking 1 ppm of additional scattering losses into account.

The distributions confirm the expected trends for both procedures, and allow us to draw two important conclusions: firstly, the electronic slew rate does not limit the ring-down measurements, since the significantly shorter photon lifetime at 1604 nm is clearly retrievable. Secondly, the finesse values stated in the main text can be taken as conservative.

**Acknowledgements:**
We thank Felix Donnerbauer for assistance during the measurements. We gratefully acknowledge the Austrian Science Fund (FWF) within project P-27297 (CaviCool) and W1210-N25 (COQUS), and I 3167-N27 (SiC-EiC), the WWTF within the project ICT12-041 (PhoCluDi), and the TU Innovative Projekte for financial support. SK acknowledges financial support from the Erwin Schrödinger Center for Quantum Science & Technology (ESQ) of the Austrian Academy of Sciences in the project ROTOQUOP. We furthermore thank R. Lalezari and R. Patterson for useful discussions.

# Supplementary Information for
# Silicon microcavity arrays with open access and a finesse of half a million

Georg Wachter[1,2], Stefan Kuhn*[,1], Stefan Minniberger[1], Cameron Salter[1], Peter Asenbaum[1], James Millen[1,+], Michael Schneider[3], Johannes Schalko[3], Ulrich Schmid[3], André Felgner[4], Dorothee Hüser[4], Markus Arndt[1], and Michael Trupke*[,1,2]

[1]Faculty of Physics, University of Vienna, VCQ, Boltzmanngasse 5, 1090 Vienna, Austria
[2]Institute for Atomic and Subatomic Physics, TU Wien, VCQ, Stadionallee 2, 1020 Vienna, Austria
[3]Institute for Sensor and Actuator Systems, TU Wien, 1040 Vienna, Austria
[4]Physikalisch-Technische Bundesanstalt, Bundesallee 100, D-38116 Braunschweig
[+]Current address: Department of Physics, King's College London, Strand, London WC2R 2LS United Kingdom

*Correspondence and requests for materials should be addressed to stefan.kuhn@univie.ac.at or michael.trupke@univie.ac.at

**Determination of losses:**

The finesse of an optical cavity is limited by several factors, including the transmission $T$ and absorption $A$ of the mirrors, scattering losses $l_S$ due to short-scale surface roughness[1], clipping losses $l_C$ caused by the finite aperture $R_O$ of the mirrors and mode distortion losses $l_M$ caused by large-scale mirror deformations.

The scattering and clipping losses can be estimated from

$$l_S = 1 - e^{-(4\pi\sigma/\lambda)^2} \cong \left(\frac{4\pi\sigma}{\lambda}\right)^2 \text{ and}$$

$$l_C = \frac{2}{\pi w_M^2} \int_0^{R_0} e^{-2r^2/w_M^2} \, 2\pi r \, dr = e^{-2R_O^2/w_M^2}.$$

The scattering losses are derived from a sinusoidal grating equation assuming white spatial noise with an rms amplitude of $\sigma$, while the clipping losses are calculated by integrating the power transmitted through an aperture of radius $R_O$. We can now place bounds on the possible reductions of the reflectivity values $\rho_{1,2}$ of the mirrors from the observed finesse,

$$F = \frac{\pi(\rho_1\rho_2)^{1/4}}{1-\sqrt{\rho_1\rho_2}} \cong \pi\left[T + A + \frac{l_{S,1}+l_{C,1}+l_{M,1}}{2} + \frac{l_{S,2}+l_{C,2}+l_{M,2}}{2}\right]^{-1}.$$

The indices pertain to the two mirrors. We have assumed that all losses are small and that the coating properties are the same for all surfaces.

*Losses due to short-scale roughness:* Values for $l_S$ were obtained from stylus and optical profilometer, as well as atomic force microscope (AFM) measurements, which were performed after the etch procedure and after oxidation polishing.

Before polishing, profiles within a circular area of a radius of 20 μm, recorded with a tactile profilometer (Tencor P17), display a typical rms roughness of around Rq = 2.5±0.8 nm.

After smoothing by oxidation, the tactile measurement delivered an upper limit of Rq < 2 nm. Optical (phase-shifting interference microscope) measurements on the polished mirrors gave

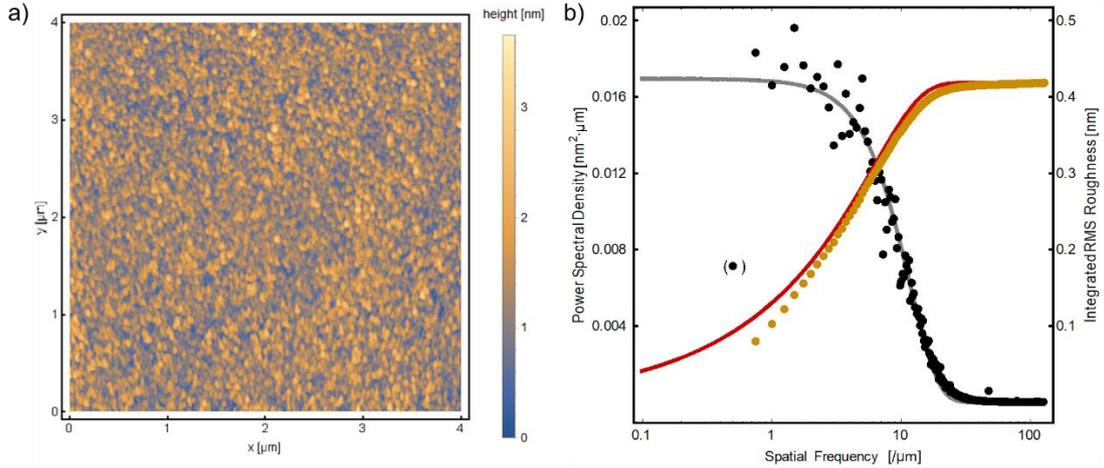

**Figure S1:** *Short-scale roughness of the silicon surface. a) AFM data after removal of spherical background and parabolic fit to each line. b) The extracted power spectral density of the roughness (black dots) is shown with a Gaussian fit (grey line). (The bracketed point was removed for the fit). Cumulated roughness of the spectral components (yellow dots) and corresponding function extrapolated from the fit (red line).*

Rq values of below (0.5±0.3) nm. The optical measurements have an evaluated profile length of 15 µm with 60 data points. AFM measurements confirm this result: An example AFM measurement, displaying a roughness of $\sigma = 0.42$ nm (rms) over a measurement area of 4×4 µm² with 1024×1024 points, is shown in Fig. S1 a). The RMS roughness value Rq was computed after removing a spherical shape and a line-by-line parabolic fit, but without further spectral filtering, from the standard deviation of the height values.

The properties of the substrate are expected to be perfectly replicated by the mirror coating[2]. This roughness value would therefore lead us to expect a loss of $l_s = 10.5$ ppm, corresponding to a maximum finesse of 2.8×10⁵. The finesse we measure in both types of device instead points to a maximal loss of $l_{max} = 2.8$ ppm, corresponding to a roughness of $\sigma_{max} = 0.21$ nm. It is therefore likely that not all roughness components retrieved in the AFM measurement contribute equally to the scattering loss.

The spatial power spectral density (PSD) of the amplitude noise (Fig. S1 b)) of the uncoated micromirrors is closely matched by a Gaussian centred at zero,

$$PSD = P_0 e^{-(f_s/f_e)^2}$$

with a $1/e$-width of $f_e = 11.66$ µm⁻¹, evaluated from the average PSD of all 1024 lines. Such a spectrum indicates a Gaussian autocovariance function[3]. The step size on the measurement was 4 nm/pixel and the tip radius was 10 nm, such that both enable the measurement of far higher spatial frequencies than the inverse of the observed correlation length.

Such a coloured spectrum limits the validity of the simple expression for $l_S$ used above and indicates the need for a refined analysis. Distortions on a markedly greater scale than the beam waist can be neglected: Pure sine components correspond to a tilt of the mirror, while pure cosine contributions only modify the curvature. Spatial frequencies $f_S$ smaller than the mode waist can be assumed to have the effects expected from the standard scattering approximation. However, surface amplitude noise with a spatial wavelength smaller than the optical wavelength can be expected to contribute only weakly to the loss of the specular component. We therefore cumulate the spectral components to find the maximum relevant spatial frequency of the surface roughness[4]. Assuming a hard cut-off, we calculated the resulting roughness with $\sigma_{RMS} = \sqrt{\sum_{f_{min}}^{f_s(max)} PSD(f_s)/f_{min}}$, with $f_{min} = 0.25$ µm⁻¹, from the data and $\sigma_{RMS} = \sqrt{\frac{\sqrt{\pi}}{2} P_0 f_e \mathrm{erf}(f_s/f_e)}$ by integration of the fit.

We find that the achieved finesse values can only be achieved if features with a spatial extent smaller than $f_s(\max) = 2.65/$µm, i.e. $1/f_s(\max) = \lambda/4.1$ do not contribute to the loss at $\lambda = 1.55$ µm. Thorough analyses of the effect of roughness on the scattering properties support this reasoning, and indicate that even smaller values of $f_s(\max)$ should be used[3]: According to the generalized Harvey-Shack surface scatter theory, since a grating with a period smaller than $\lambda$ will not scatter light into modes with real-valued amplitude, the spectral portion $f_s > 1/\lambda$ should be excluded. This limit yields a relevant surface roughness of $\sigma_{rel} = 0.10$ nm, enabling a theoretical maximum finesse of 4.8×10⁶.

*Losses due to shape distortions:* The mirror diameters range from 60 µm $< 2R_O <$ 120 µm, making pure clipping losses negligible since $w_M < 10$ µm for all measured cavities. Assuming a loss of $l_{max} = 2.8$ ppm, the effective aperture of the mirrors is on the order of 2.53 $w_M$ indicating that the mirror size is not a relevant constraint. However, light scattering into higher-order modes can be related to an effective aperture of the mirrors. Since the losses due to generic mirror shape distortions cannot be calculated analytically, we determine a qualitative factor to provide insight into the characteristics of the mirrors. The ideal phase front of a Gaussian beam is a pure parabola and the lowest symmetric distortion is given by a quartic term. The additional phase shift as a function of the radial coordinate $r$ is given by $\Delta\phi(r) = 2Sr^4$. Numerical simulations[5] indicate that a value of $S < 10^{-7}$ is necessary in order to reach a finesse $F > 5 \times 10^5$ in the symmetric cavity assembly. Distortions on this scale are challenging to quantify, since they require a local height resolution of nanometres on a parabolic background with a size of several micrometres. Nonetheless, the calculated limiting value of $S$ shows that the mirrors must match the desired parabolic profile to an extremely high degree.

**Birefringence:**
The low birefringence observed in most of our microcavities indicates that the mirror shapes must be highly symmetric. The observed frequency splitting of the orthogonal polarization states could be related to mirror imperfections caused by the etching process, but also due to mirror misalignment or tilt. It was previously shown[6] that a difference in radii of curvature (along two orthogonal axes) can lead to such a splitting, where the length-independent phase shift is given by

$$\delta\phi = \frac{1}{k}\frac{1-R_X/R_Y}{R_X}.$$

For a symmetric CC cavity with a radius of curvature along one azimuthal axis $R_X = 201$ µm, and assuming equal shifts on both surfaces, this expression gives an orthogonal radius $R_Y = 202.9$ µm. The observation of birefringence does not necessarily require an asymmetry of the shape itself, but can also be caused by an off-axis displacement of the cavity mode spot: The local radius of curvature of a parabola with $z = (x^2 + y^2)/2R(0)$ is given by $R_Y(y) = R \times \left[1 + \frac{y^2}{R(0)^2}\right]^{3/2}$. The orthogonal radius of curvature increases more slowly, with $R_X(y) = R(0) \times \left[1 + \frac{y^2}{R(0)^2}\right]^{1/2}$.

For perfectly parabolic mirrors, an off-axis displacement of their centres of 19.5 µm, corresponding to a tilt of the cavity axis by 15°, would be required to cause such a splitting. This displacement or tilt is more than an order of magnitude greater than expected from our fabrication tolerances, and is therefore not expected to be the dominant contribution to the observed splitting. Aside from assembly effects, strain in the coating can contribute to the birefringence at the level of several µrad[7].

Finally, the strong correlation between finesse and birefringence (see main text, Fig. 3 a)) indicates a common cause: It is likely that both arise from residual shape deviations with a spatial wavelength on the order of the mode size, since such distortions can lead to scattering into higher-order modes and to a small angular dependence of the radius of curvature[5,6].